\documentclass[aps,prl,reprint,amsmath,amssymb,longbibliography]{revtex4-2}
\usepackage{graphicx}
\usepackage{bm}
\usepackage[colorlinks=true,allcolors=blue]{hyperref}
\makeatletter
\@booleantrue\acknowledgments@sw
\makeatother
\newcommand{\ii}{\mathrm i}
\newcommand{\Tr}{\operatorname{Tr}}
\newcommand{\Li}{\operatorname{Li}}
\newcommand{\heading}[1]{\par\smallskip\noindent\textit{#1.---}}
\newcommand{\cV}{\mathcal V}
\newcommand{\cI}{\mathcal I}
\newcommand{\nh}{\widehat N}
\newcommand{\nn}{\mathfrak n}
\newcommand{\ttop}{\mathfrak t}
\begin{document}
\title{The chiral $M^{3,2}$ theory: partition functions beyond large-$N$ and quantum holography}
\author{Xuao Zhang}
\email{zhangxuao@ucas.ac.cn}
\affiliation{University of Chinese Academy of Sciences, Beijing 100190, China}
\date{November 17, 2026}
\begin{abstract}
 We identify families of Bethe vacua in the chiral \(M^{3,2}\) theory whose individual free energies grow as \(N^{3/2}\). High-precision data support closed forms for their all-order perturbative \(1/N\) expansions. At a distinguished mixed Chern--Simons coupling, the baryonic magnetic flux distinguishes members of this family of vacua. We also study the three-sphere partition function numerically in the 't Hooft limit and Cardy limit, and propose an all-orders Airy formula. The four-derivative and one-loop corrections are precision tests of holography.
\end{abstract}
\maketitle

\section{Introduction}
For $N$ M2-branes probing the Calabi-Yau fourfold cone $X_8 = C(Y_7)$, where $Y_7$ is a Sasaki--Einstein manifold, the AdS/CFT correspondence defines the M-theory path integral in ${\rm AdS}_4 \times Y_7$ by the supersymmetric gauge theory on its boundary. Supersymmetric localization reduces the field theory path integral to an ordinary integral, enabling black-hole microstate counting \cite{Benini:2015eyy} and precision tests beyond the leading large-$N$ limit \cite{Bobev:2022jte,Bobev:2022eus,Bobev:2023lkx,Bobev:2022wem,Bobev:2024mqw,Bobev:2025ltz}.

We are interested in three specific field theory partition functions \cite{Closset:2019hyt} amenable to supersymmetric localization: the topologically twisted index (TTI), which counts BPS states of a magnetic background; the superconformal index (SCI), which counts BPS states of a Kerr--Newman black hole; and the squashed-sphere partition function. Holomorphic factorization \cite{Beem:2012mb} indicates an underlying relation among them, which is manifest in the Cardy limit \cite{Choi:2019dfu} and leads to the OSV-type relations \cite{Bobev:2026lvl}. These relations offer new insights into holography and ${\rm AdS}_4$ black holes.

For $M^{3,2}$, also known as $M^{1,1,1}$, the proposed gauge theory is a chiral quiver~\cite{Martelli:2008si,Hanany:2008fj,Benini:2011cma} whose long-range eigenvalue forces obstruct the usual large-$N$ analysis. Symmetrizing the sphere matrix model or counting chiral operators reproduces the leading $S^3$ free energy predicted by gravity~\cite{Amariti:2011uw,Gang:2011jj,Gulotta:2011aa,Kim:2019umc}. With suitable mixed Chern--Simons couplings, the low-energy SCI agrees with the supergravity spectrum \cite{Cheon:2011refined,Gang:2011jj}. Assuming that the large-\(N\) relation \cite{Hosseini:2016tor} between the TTI and sphere free energy extends to chiral quivers yields a leading TTI consistent with gravity~\cite{Kim:2019umc,Hosseini:2019geometry}. The geometry also predicts black-hole entropies with baryonic charge~\cite{Hosseini:2025master} and supplies data for an Airy-function proposal on the squashed sphere~\cite{Cassia:2025jkr} (with which we find disagreement, see below). Despite this progress, how the expected \(N^{3/2}\) free energy emerges from the original \(M^{3,2}/\mathbb{Z}_k\) matrix model has remained unclear.

For chosen mixed-CS levels, we numerically obtain all Bethe roots at finite $N$ and follow them to larger $N$. Recent work starts from the saddle-point equations for the sphere free energy~\cite{Hosseini:2025jxb}. We start from the Bethe equations for the TTI, which are polynomial. This allows us to apply homotopy continuation to obtain a large set of solutions. We obtain seed solutions by applying Newton's method to random initial guesses. We then track these solutions as the parameters vary along closed paths, which can reveal additional roots of the original equations. In this way, we can find almost all the Bethe roots with the expected large-$N$ behavior. Based on the Bethe roots, we also find a solution of the sphere saddle-point equations, which gives a precise, independent numerical test of the OSV-type relations. 

The mixed-CS levels affect the solutions through the parameter $D$ defined in \eqref{eq:em-parameters}. In this draft, we focus on $D = 9$, where the structure of Bethe roots is especially clear: using the finite-$N$ solutions, we can identify all Bethe roots with the expected $O(N)^{3/2}$ scaling.

\section{The theory and the Bethe Ansatz formula}
\begin{figure}[!htb]
	\centering
	\includegraphics[width=0.70\columnwidth]{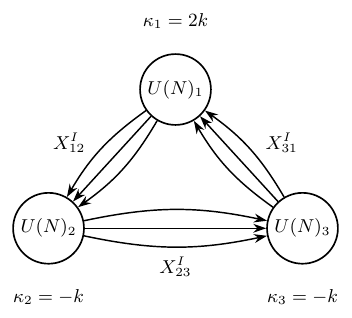}
	\caption{The chiral $M^{3,2}$ quiver with ordinary Chern--Simons levels $(2k,-k,-k)$. Each set of three arrows represents the matter fields $X^I_{ab}$, $I=1,2,3$.}
	\label{fig:quiver}
\end{figure}
The matter content is summarized in Fig.~\ref{fig:quiver}. The gauge group is $U(N)_1\times U(N)_2\times U(N)_3$, with ordinary Chern--Simons (CS) levels $\kappa_a =(2k,-k,-k),\, a=1,2,3$. Three fields $X^I_{ab}$, labeled by flavor $I=1,2,3$, transform as $(\overline{\mathbf N}_a,\mathbf N_b)$ on each edge $(a,b)=(1,2),(2,3),(3,1)$, with superpotential $W=h_W \epsilon_{IJK}\Tr(X^K_{31}X^J_{23}X^I_{12})$. Mixed CS terms couple the three trace gauge fields. In this work, we choose $\Lambda_{aa}=3$, $\Lambda_{a\ne b}=-3/2$, which satisfy the parity-anomaly cancellation conditions. We work at the symmetric mesonic point with chemical potentials \(\Delta_I=2\pi/3\) of the universal twist \cite{Azzurli:2017-universal, Bobev:2017uzs} and leave baryonic mixing with the R-symmetry for future work. 

The TTI is a supersymmetric partition function on $\Sigma_{\mathfrak{g}} \times S^1$, where the R-symmetry gauge field gives a topological twist on $\Sigma_{\mathfrak{g}}$. We introduce the fugacities and holonomies $x_i^{(a)}=e^{\ii u_i^{(a)}}$, $P_a=\prod_i x_i^{(a)}$, and $y=e^{2\pi\ii/3}$. The poles that contribute to the index are solutions to the Bethe Ansatz equations (BAE):
\begin{equation}
\mathcal B_i^{(a)}=-\bigl(x_i^{(a)}\bigr)^{\kappa_a}
 \left(\frac{P_a}{P_b}\right)^3
 \prod_j\left(\frac{1-yx_j^{(b)}/x_i^{(a)}}
 {1-y^{-1}x_j^{(c)}/x_i^{(a)}}\right)^3=1,
\label{eq:bae}
\end{equation}
where $(a,b,c)$ is cyclic. The topological fugacities are chosen to be $\xi_a=(-1)^{N+1}$.

We focus on genus zero. The extension to $\mathfrak{g}>1$ is straightforward.\footnote{For a given Bethe root, its free energy at genus $\mathfrak{g}$ and genus zero are related by $F_{\mathfrak{g}}(\Delta, \mathfrak{n}) = (1-\mathfrak{g}) F_{0}\left(\Delta, \frac{\mathfrak{n}}{1-\mathfrak{g}}\right)$~\cite{Benini:2016hjo}.} At genus zero, localization gives \cite{Benini:2015noa,Benini:2016hjo,Closset:2016arn}
\begin{equation}
\begin{gathered}
Z_{\rm TTI}=\sum_\alpha Z_\alpha,\qquad
 J=\frac{\partial\log\mathcal B}{\partial(\ii u)},\\
 Z_\alpha=\left.\frac{\mathcal D_{\rm vec}
 \prod_I\mathcal D_I^{1-\nn_I}\prod_aP_a^{\ttop_a}}
 {\det J}\right|_{u_\alpha}.
\end{gathered}
\label{eq:residue}
\end{equation}
The sum is over regular Bethe vacua modulo eigenvalue permutations. Here $\mathcal D_{\rm vec}=\prod_{a,i\ne j}(1-x_i^{(a)}/x_j^{(a)})$ and $\mathcal D_I=e^{3\ii N^2\Delta_I/2}/\prod_{(a,b),i,j}(1-e^{\ii\Delta_I}x_j^{(b)}/x_i^{(a)})$. We use $\nn_I =(0,1,1)$ and initially $\ttop_a=0$. Equal $\Delta_I$ make all $\mathcal D_I$ equal, so redistributing the chiral fluxes at fixed $\sum_I\nn_I=2$ leaves this expression unchanged.

The SCI is defined on $S^2 \times_\omega S^1$ with rotational fugacity $e^{\ii\pi\omega}$. In its Cardy limit $\omega\to\ii0^+$, the same roots contribute \cite{Choi:2019dfu, Bobev:2022wem, Bobev:2024mqw}:
\begin{equation}
\log\cI_\alpha=-\frac{\operatorname{Im}\cV_\alpha}{\pi\omega}
 +\log|Z_\alpha|+\ii\varphi_\alpha +O(\omega).
\label{eq:cardy}
\end{equation}
The Bethe potential is given by
\begin{equation}
\cV_\alpha=-\tfrac12u^TKu-t\cdot S
 +\sum_{(a,b),i,j,I}f(u_j^{(b)}-u_i^{(a)}+\Delta_I)
 +2\pi\ell\cdot u,
\label{eq:potential}
\end{equation}
with $f(z)=\Li_2(e^{\ii z})-z^2/4$, $S_a=\sum_i u_i^{(a)}$, $K_{ai,bj}=\kappa_a\delta_{ab}\delta_{ij}+\Lambda_{ab}$, and $e^{\ii t_a}=\xi_a$. The integer vector $\ell$ selects a stationary sheet. \(\varphi_\alpha \) denotes an overall phase.

\section{The large-$N$ numerics}

\begin{figure}[!htb]
\centering
\includegraphics[width=\columnwidth]{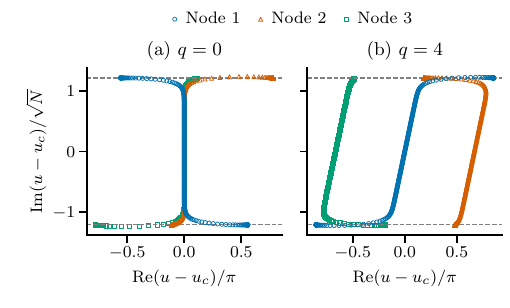}
\caption{Numerical Bethe roots at $N=299$ for (a) the equal-phase $q=0$ and (b) the clock-phase $q=4$ families. Colors and symbols distinguish nodes. The imaginary coordinate is scaled by $\sqrt{N}$. Dashed lines mark the limiting endpoints $\pm T$. }
\label{fig:roots}
\end{figure}

We use homotopy continuation to find Bethe roots at finite $N$~\cite{Peng:2026udd} and track them to higher rank with high precision. Figure~\ref{fig:roots} shows two examples.

\subsection{Structure of the roots}

Focusing on $k = 1$, we identify families of Bethe roots labeled by an integer $q = 0, 1\mod{3}$. For the two remainders of $q$, the central phases of the roots coincide or form a three-step clock, respectively, as illustrated in Fig.~\ref{fig:roots}. In both cases, a branch consists of a broad linear core region which joins two sharply clustered tails through curved transition regions.

The leading imaginary distribution is common to these families. We introduce $t$ to parameterize the rescaled imaginary part of $u_i^{(a)}$ in the core region:
\begin{equation} \label{eq:ua}
	u^{(a)}=u_c+\ii\sqrt N t+a_v t+\phi_a+\ii\zeta_a/\sqrt N+\cdots,
\end{equation}
where $u_c$ is the central root of node 1. The sign of the common slope $a_v = \pm \frac{\sqrt{3}q}{18}$ depends on the remainder of $q$ modulo three. The relative phases $\phi_a$ vanish for the equal-phase core and take the values $(0,2\pi/3,-2\pi/3)$ for the clock core. The special case is $q = 0$, for which the core is vertical in the $u$-plane.

\begin{figure}[!htbp]
\centering
\includegraphics[width=\columnwidth]{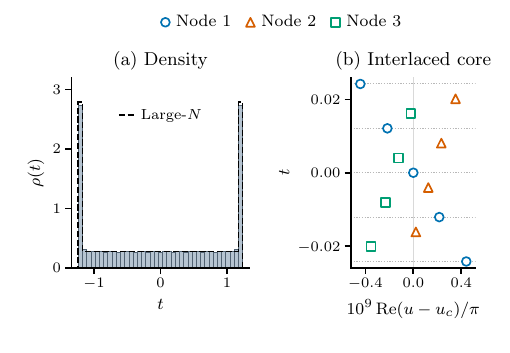}
\caption{Core interlacing and density for the $D=9$, $q=0$ vacuum at $N=299$. (a) Histogram of all $3N$ roots, normalized by $3N\,\Delta t$ for bin width $\Delta t$. The dashed curve is Eq.~\eqref{eq:density} averaged over the same bins. (b) Roots near the center $u_c = 0$. Successive roots cycle through nodes $1,3,2$ in increasing $t$.}
\label{fig:core-density}
\end{figure}

As shown in Fig.~\ref{fig:core-density}, the distribution of $t \equiv {\rm Im}(u)/N^{1/2}$ has density
\begin{equation}
	\begin{aligned}
		\rho(t) &= \rho_c \mathbf1_{(-T,T)}
		+\frac16[\delta(t-T)+\delta(t+T)],\\ \rho_c &= \frac{\sqrt3}{2\pi},\quad 
		T=\frac{2\pi}{3\sqrt3}. \\
	\end{aligned}
\label{eq:density}
\end{equation}
Two thirds of the roots occupy the core, and one sixth accumulates at each tail. Chiral-operator counting predicted the same normalized density in the sphere-partition function \cite{Gulotta:2011aa}, we further resolve the internal structure of the tail clusters: distances $\epsilon$ from the boundary are exponentially small, $|\epsilon|=e^{-\sqrt N Y}$. Its logarithm remains of order $\sqrt N$, so the tail contributes to the leading free energy despite collapsing to a point in~\eqref{eq:density}. The tail distribution in $Y$ has density $\sigma(Y)=9(T-Y)/(4\pi^2)$ for $0<Y<T$. \footnote{ The upper bounds of $t$ and $Y$ coincide at $T$ for this choice of mixed CS levels. They can differ at other levels. }

The residual long-range force plays an important role in the large-$N$ BAE. Summing the small relative displacements $\zeta_a$ in \eqref{eq:ua} over $N$ roots produces an order-$\sqrt N$ force that balances the force from the CS terms. The large-$N$ BAE give
\begin{equation}
\kappa_a+3\rho_c(\zeta_b-\zeta_c)=0,\qquad
(\zeta_1,\zeta_2,\zeta_3)=\frac{(0,-1,1)}{3\rho_c}.
\label{eq:force}
\end{equation}
This structure is consistent with the numerical results in Fig.~\ref{fig:core-density}. Core--tail matching and normalization fix $T$, and a finite Bethe branch integer fixes the remaining tilt $a_v$. A detailed treatment of the large-$N$ BAE will be reported later \cite{Zhang:M32toappear}.

\subsection{Closed-form expressions for the indices}
We continue the families labeled by $q$ at $k=1$ to generic $k$. We introduce $Q_q(k)=q$ for equal-phase roots and $Q_q(k)=q-2(k-1)$ for clock-phase roots. With $F_q(N,k)=-\log|Z_q^{\rm TTI}(N,k)|$ and $a(k)=16\sqrt{k}\pi/(9\sqrt3)$, we find strong numerical evidence for the closed-form expression
\begin{equation}
\begin{aligned}
F_q^{\rm pert}(N,k)={}&a(k)\left[\nh_q(k)^{3/2}
 +\mathfrak c(k)\nh_q(k)^{1/2}\right]\\
 &+\frac12\log\nh_q(k) - \hat{f}_0(k),
\end{aligned}
\label{eq:tti}
\end{equation}
with shifted membrane charge and relative coefficient
\begin{equation}
	\begin{aligned}
		\nh_q(k)&=N+\frac{7k}{36}+\frac1{4k}-\frac{Q_q(k)^2}{108k},\\
		\mathfrak c(k)&=\frac1k\left(\frac{Q_q(k)^2}{72}-\frac34\right).
	\end{aligned}
\end{equation}
The constant is
\begin{equation}
\begin{aligned}
- \hat{f}_0={}&9\mathcal A_{\rm ABJM}(2k)-\frac{27\zeta(3)}{2\pi^2}k^2\\
 &+\frac32\log k+\log(6\sqrt3),
\end{aligned}
\label{eq:kconstant}
\end{equation}
where $\mathcal A_{\rm ABJM}$ is the ABJM constant-map function \cite{Bobev:2022jte}. In the large-$k$ expansion of $\hat{f}_0(k)$, the $\log k$ term exactly cancels. Expanding the shifted rank gives
\begin{equation}
\begin{aligned}
F_q^{\rm pert}(N,k)={}&a(k)N^{3/2}+a(k)\frac{7k^2-9}{24k}N^{1/2}\\
 &+\frac12\log N - \hat{f}_0(k)+O(N^{-1/2}).
\end{aligned}
\label{eq:shared}
\end{equation}
The leading coefficient equals the gravitational prediction. At fixed $k$, the first correction and logarithm are also independent of $q$; different $q$ enter only at $N^{-1/2}$. For $q\in3\mathbb{N}$, the $q$ and $-q$ branches share a perturbative formula. Their roots and complex phases can still differ.

\begin{figure}[t]
\centering
\includegraphics[width=\columnwidth]{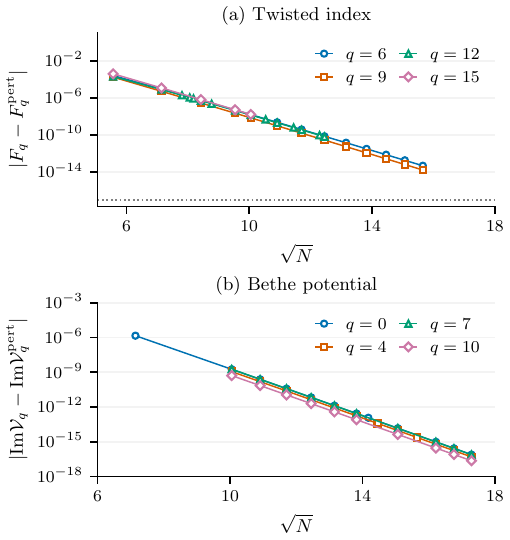}
\caption{Numerical differences for $k=1$ Bethe roots. (a) Absolute differences between single-vacuum TTI and Eq.~\eqref{eq:tti}; the dotted line marks the uncertainty in $C$. (b) Absolute differences between Bethe potentials and Eq.~\eqref{eq:vresult}.}
\label{fig:errors}
\end{figure}

The high-precision data also support
\begin{equation}
	\frac{\operatorname{Im}\cV_q^{\rm pert}(N,k)}{2\pi}
	=  \frac{ a(k) s_q }{4}\nh_q(k)^{3/2} + \hat{g}_0(k),
	\label{eq:vresult}
\end{equation}
where
\begin{equation}
	 s_q=(-1)^{q\, {\rm mod}\, 3}, \qquad  \hat{g}_0(k) = - s_q \frac{9k^2\zeta(3)}{2\pi^2}.
	 \label{eq:g0}
\end{equation}
 Fig.~\ref{fig:errors} shows the exponential decay of the deviations from both closed expressions at $k=1$, consistent with nonperturbative instanton corrections. Rank-independent terms in Bethe observables have also been reconstructed for other M2-brane theories \cite{Hong:2026zul, Hosseini:2026dkj, Hosseini:2026lvj, Hosseini:2026kyx}.

Eqs.~\eqref{eq:cardy}, \eqref{eq:tti}, and~\eqref{eq:vresult} determine the first two orders of the Cardy expansion of the SCI, to all perturbative orders in $1/N$:
\begin{equation}
\begin{aligned}
-\log\cI_q^{\rm pert}(N,k)={}&\frac{s_qa(k)}{2\omega}\nh_q(k)^{3/2}
 -\frac{9s_qk^2\zeta(3)}{\pi^2\omega}\\
 &+F_q^{\rm pert}(N,k)-\ii\varphi_q+O(\omega).
\end{aligned}
\label{eq:sci}
\end{equation}
The $+\frac12\log N$ in $-\log Z$ agrees with the universal genus-zero gravitational logarithm \cite{Liu:2017vbl,PandoZayas:2020iqr,Bobev:2023dwx}. This quantum test is sharper than matching the classical $N^{3/2}$ power. 

General topological fluxes $\ttop_a$ multiply each vacuum contribution by $\prod_a P_a^{\ttop_a}$~\cite{Closset:2017zgf}, shifting its free energy by $\Delta F_q=\sum_a\ttop_a\operatorname{Im}S_a$. For the reflection-symmetric roots studied here, $\operatorname{Im}S_1=0$ and the product of the Bethe equations gives $\operatorname{Im}S_2=-\operatorname{Im}S_3$. The magnitude of each contribution thus depends only on the baryonic flux combination $\ttop_2-\ttop_3$. \cite{Klebanov:2010-betti} Using the numerical roots, we find
\begin{equation}
\begin{aligned}
\Delta F_q&=-\frac{\ttop_2-\ttop_3}{2}\operatorname{Im}(S_3-S_2)\\
 &\simeq-(\ttop_2-\ttop_3)\frac{\pi Q_q(k)}{27\sqrt{3k}}\sqrt{\nh_q(k)}.
\end{aligned}
\label{eq:flux}
\end{equation}
This distinguishes roots with different $Q_q(k)$. The bulk interpretation of this branch-dependent response to baryonic magnetic flux remains open.

\section{Four-derivative corrections, Airy conjecture, and OSV relations}

The $q=0$ Bethe root is special: it is stationary under baryonic chemical potential, and its slope vanishes, as does that of the sphere saddle. These observations motivate identifying it with the universal sector \cite{Azzurli:2017-universal}.

Consistent truncation of 11d supergravity on the internal space $M^{3,2}$ gives four-dimensional gauged supergravity \cite{Cassani:2012truncation}. In the framework of 4d $\mathcal{N} = 2$ gauged supergravity, the first correction to the Gibbons-Hawking action comes from curvature-squared interactions. Matching the $N^{1/2}$ terms of the TTI and SCI fixes the two combinations entering its minimal-sector on-shell action \cite{Bobev:2020egg,Bobev:2021oku}:
\begin{equation}
\begin{aligned}
I_{4\partial}(\mathcal M)=a(k)\bigg[&\mathcal F N^{3/2}\\
 &+\left(\frac{7k^2+9}{24k}\mathcal F-\frac3{8k}\chi\right)N^{1/2}
\bigg]+o(N^{1/2}).
\end{aligned}
\label{eq:gravity}
\end{equation}
Here $\mathcal F$ is the normalized classical action, and $\chi$ is the Euler characteristic. This expression applies to every background $\mathcal{M}$.

For a squashed sphere with $Q=b+b^{-1}$, its bulk has $\mathcal F=Q^2/4$ and $\chi=1$. Assuming the proposed Airy form \cite{Bobev:2025ltz} and identifying the \(q=0\) vacuum with the universal sector, we use the index data in Eqs.~\eqref{eq:tti} and~\eqref{eq:sci} to fix the squashed-sphere coefficients \(\mathcal B\) and \(\mathcal C\) that holds for general $b$:
\begin{gather}
Z_{S_b^3}^{\rm pert}(k)=e^{\mathcal A(b,k)}\mathcal C^{-1/3}
 \operatorname{Ai}[\mathcal C^{-1/3}(N-\mathcal B)],\label{eq:airy}\\
\mathcal C=\frac{27}{4\pi^2kQ^4},\qquad
\mathcal B=-\frac{7k}{36}-\frac1{4k}+\frac1{kQ^2}.\label{eq:airycoeff}
\end{gather}
 Our precise result disagrees with the \(M^{1,1,1}\) formula of \cite{Cassia:2025jkr}, which gives \(\mathcal B=Q^{-2}-2/9\) at $k=1$. The OSV-type relations below determine the small-$b$ expansion of $\operatorname{Re}\mathcal A(b,k)$ through order $b^0$, as given in \eqref{eq:airy-normalization-cardy}.

The OSV-type relations join two sphere contributions to form an index \cite{Ooguri:2004-OSV,Bobev:2026lvl}. With $U_\pm=(1\pm\omega)^2/\omega$, $\rho_\pm=(4\pi^2kU_\pm^2/27)^{1/3}$, and $X_\pm=\nh_0(k)\mp(kU_\pm)^{-1}$, these relations read
\begin{align*}
\cI_{{\rm SCI},0}^{\rm pert}(\omega;k)&\simeq
\mathcal N_{\rm SCI}\rho_+^2\operatorname{Ai}(\rho_+X_+)^2,\\
Z_{{\rm TTI},0}^{\rm pert}(\omega;k)&\simeq
\mathcal N_{\rm TTI}\rho_+\zeta_A\rho_-
\operatorname{Ai}(\rho_+X_+)\operatorname{Ai}(\zeta_A\rho_-X_-),
\end{align*}
with $\zeta_A=e^{\pm2\pi\ii/3}$ from lateral continuation of the second sphere factor. This rotation cancels the $1/\omega$ poles of the Airy exponents in the TTI. In the Cardy limit $\omega\to0^+$, matching the SCI to Eq.~\eqref{eq:sci} and the unrefined TTI to Eq.~\eqref{eq:tti} gives
\begin{equation}
\begin{aligned}
\operatorname{Re}\log\mathcal N_{\rm SCI}
 &=-\frac{2\hat g_0(k)}{\omega}+\log\!\left(\omega\sqrt{\frac{108}{k}}\right)\\
 &\quad+\hat f_0(k)+O(\omega).\\
\end{aligned}
\label{eq:index-normalizations-cardy}
\end{equation}
\begin{equation}
\operatorname{Re}\log\mathcal N_{\rm TTI}
=\log\!\left(\omega\sqrt{\frac{108}{k}}\right)+\hat f_0(k)+O(\omega).
	\label{eq:airy-normalization-cardy-tti}
\end{equation}
The OSV-type relation identifies $\operatorname{Re}\log\mathcal N_{\rm SCI}$ with the sum of $\operatorname{Re}\mathcal A$ for the two sphere saddles at $b^2=\omega$. Equation~\eqref{eq:index-normalizations-cardy} then gives
\begin{equation}
\begin{aligned}
\operatorname{Re}\mathcal A(b,k)
 &=-\frac{\hat g_0(k)}{b^2}+\log b\\
 &\quad+\frac{\hat f_0(k)}{2}+\frac14\log\frac{108}{k}+O(b^2),
\end{aligned}
\label{eq:airy-normalization-cardy}
\end{equation}
with $\hat{f}_0$ in \eqref{eq:kconstant} and $\hat{g}_0$ in \eqref{eq:g0}.

\section{Squashed sphere partition function}
In this section, we give independent strong evidence supporting the Airy conjecture \eqref{eq:airy}. 

\subsection{The 't Hooft limit with no squashing}

\begin{figure}[t]
	\includegraphics[width=\columnwidth]{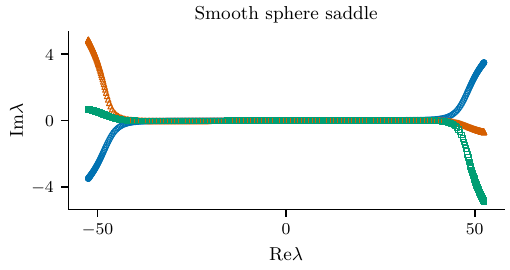}
	\caption{ The smooth and connected sphere saddle at $N=401$, with complex eigenvalues; node symbols match Fig.~\ref{fig:roots}. }
	\label{fig:sphere}
\end{figure}

For vanishing masses and Fayet--Iliopoulos sources, and a common trial R-charge $r=2/3$ for all chiral multiplets, the round-sphere partition function is \cite{Jafferis:2010-Zextremization,Hosseini:2025jxb}
\begin{equation}
	\begin{aligned}
		Z_{S^3}&=\frac1{(N!)^3}\int\prod_{ai}\frac{d\lambda_{ai}}{2\pi}\,
		e^{\ii\lambda^TK\lambda/(4\pi)+3\sum_{a,i,j}\ell(z_{aij})}\\
		&\hspace{1cm}\times\prod_{a,i<j}4\sinh^2\!\frac{\lambda_{ai}-\lambda_{aj}}2,
	\end{aligned}
	\label{eq:sphere}
\end{equation}
where $z_{aij}=\frac13+\frac{\ii}{2\pi}(\lambda_{a+1,j}-\lambda_{ai})$, $\ell'(z)=-\pi z\cot(\pi z)$, and $\ell(0)=0$; \(K\) is the CS matrix defined below \eqref{eq:potential}. Denoting the integrand by $e^{-\mathcal S}$, we solve $\partial\mathcal S/\partial\lambda_{ai}=0$ for all complex eigenvalues. Motivated by the large-$N$ relation between the saddles and Bethe roots \cite{Hosseini:2016tor}, we use Bethe roots as initial guesses and then apply Newton correction and rank continuation. At $k=1$, we find a smooth single-cut saddle, seeded by $q=0$ (Fig.~\ref{fig:sphere}). \footnote{We also find other large-$N$ saddles, but each has outlying components, has discontinuities, or yields a large-$N$ coefficient that differs from the bulk prediction. Whether these saddles contribute to the integral remains open.} 
We also continue it in $k$ to probe the 't Hooft limit, where $N,k\to\infty$ with $\lambda=N/k$ fixed. Following Refs.~\cite{Geukens:2024zmt, Bobev:2025ltz}, we first fit the planar free energy $F_0$ in the genus expansion: \footnote{Note that this fit needs to include $N\log N$ and $\log N$ corrections. } 
\begin{equation}
	F_{S^3} = \sum_{{\tt g}} F_{\tt g}(\lambda) N^{2-2{\tt g}}.
\end{equation}
Then we focus on the $\lambda$-dependence of the planar free energy $F_0$. Up to corrections of order $e^{-\# \sqrt{\lambda}}$, we find strong numerical evidence for the following expression:
\begin{equation}
	F_{0,\rm pert} =\frac{1}{\lambda^2} \left[ \frac{16\pi}{9\sqrt{3}} \left(\lambda +\frac{7}{36}\right)^{3/2} - \frac{18\zeta(3)}{\pi^2} \right].
	\label{eq:sphereplanar}
\end{equation}
The shift and constant are consistent with the Airy conjecture. 

\subsection{The extremal squashing limit}
In the Cardy limit $b\rightarrow 0$, the partition function simplifies as the double-sine function appearing in the original localisation integral \cite{Hama:2011ea} admits a small-$b$ expansion. \cite{Bobev:2025ltz} The expansion of constant map \eqref{eq:airy-normalization-cardy} can be numerically reproduced by taking the difference between the full numerical integral and the Airy function. We get not only a full consistency, but also further subleading terms: denoting $\operatorname{Re}\mathcal A(b,k) = \log b + \sum_i \mathcal{A}_{(i)} b^{i}$, one has
\begin{equation}
	\begin{aligned}
		\mathcal{A}_{(2)} &= -9k \mathcal{A}_{\rm ABJM}'(2k) - \frac{7}{4},\\
		\mathcal{A}_{(4)} &= -9k^2\,\mathcal A_{\rm ABJM}''(2k) + \frac{9k}{2}\,\mathcal A_{\rm ABJM}'(2k) + \frac54 .
	\end{aligned}
\end{equation}

\section{Outlook}

One extension is to vary the chemical potentials. For the $q = 0$ root, the peaks in the density function \eqref{eq:density} are resolved. Introducing $\widehat{\Delta}_I \equiv \Delta_I/\pi$ and
\begin{equation}
	S=\sum_{I<J}\widehat\Delta_I\widehat\Delta_J,\quad
	P=\prod_{I=1}^3\widehat\Delta_I,\quad \nh \equiv N + \frac{7k}{36} + \frac{S}{18kP},
\end{equation}
we find numerically
\begin{equation}
	\begin{aligned}
		\frac{{\rm Im}\, \cV_0}{2\pi} &\simeq  \frac{\pi \sqrt{k} P}{\sqrt{S}} \nh^{3/2} + \hat{g}_0(k, \widehat{\Delta}) ,\\
	-\log|Z_0^{\rm TTI}|
		&\simeq A_k\widehat N^{3/2}-B_k\widehat N^{1/2}
		+\tfrac12\log\widehat N - \hat{f}_0(k,\widehat\Delta,\mathfrak{n}), \\
	\end{aligned}
\end{equation}
	up to exponentially suppressed corrections, with
\begin{equation}
	\begin{aligned}
		A_k
		&=\frac{\pi\sqrt{k}\,P}{\sqrt S}
		\sum_{I=1}^{3}\frac{\nn_I}{\widehat\Delta_I}
		\left(1+\frac{P}{S\widehat\Delta_I}\right),\\[4pt]
		B_k
		&=\frac{\pi}{36\sqrt{kS}}
		\sum_{I=1}^{3}\nn_I
		\left(4-3S+\frac{2S^2}{P}
		+\frac{6P}{\widehat\Delta_I^2}\right).
	\end{aligned}
\end{equation}
The leading term in the TTI agrees exactly with Ref.~\cite{Kim:2019umc}. A complete survey, as carried out in Ref.~\cite{Peng:2026udd} remains open.

Another direction is to vary the mixed CS levels so that the parameter $D$ in Eq.~\eqref{eq:em-parameters} differs from the value $D=9$ studied above. For general $D \ne 9$ and $k=1$, we identify Bethe roots with the following genus-zero free energy:
\begin{equation*}
	- \log |Z_{\rm TTI}^{\rm pert}| = a' \left[ \nh^{3/2} + \mathfrak{c} \nh^{1/2} + \mathfrak{d} \nh^{-1/2}  \right] + \log \nh - \hat{f}_0,
\end{equation*}
	with the additional $\nh^{-1/2}$ term. A remaining puzzle is that the logarithmic term differs from the bulk prediction \cite{Liu:2017vbl} by $(1-\mathfrak{g})/2$. We also identify Bethe roots whose imaginary parts scale as $O(N^{2/3})$ or $O(N)$, suggesting additional heavy sectors with an open gravitational interpretation. 
 
  Our Airy formula \eqref{eq:airy} predicts the bulk grand potential:
   \[
   	J_{\mathrm{pert}}^{M^{3,2}}(\mu;b,k)
   	= 
   	\frac{9\, \mu^3}{4\pi^2 kQ^4}
   	+\left(-\frac{7k}{36}-\frac{1}{4k}+\frac{1}{kQ^2}\right)\mu
   	+\mathcal A,
   \]
   with $\mathcal{A}$ in \eqref{eq:airy-normalization-cardy}.  From the bulk side, independent calculations using equivariant localization or the \(\mu\)-ensemble could test our results beyond leading order~\cite{BenettiGenolini:2026qdm,BenettiGenolini:2026cyc,Gautason:2025plx,Bobev:2026gir}. 

\begin{acknowledgments}
I thank Nikolay Bobev, Zhongwu Chen, and Ziruo Zhang for discussions, and especially Nikolay Bobev for comments on the draft. I would like to thank Nikolay Bobev, Pieter-Jan de Smet, Junho Hong, and Valentin Reys for collaborations on related projects. This work is supported by NSFC grant No. 12675092, 12175237.

\end{acknowledgments}

\bibliography{apssamp}

\clearpage
\twocolumngrid
\section*{End Matter}
\appendix
\section{Mixed Chern--Simons levels}
\label{app:mixed-cs}
\setcounter{equation}{0}
\renewcommand{\theequation}{A\arabic{equation}}
\renewcommand{\theHequation}{endmatter.A\arabic{equation}}

\heading{Levels and relative node directions}
The ordinary Chern–Simons levels \((2k,-k,-k)\) allow additional couplings among the three \(U(1)\) gauge fields \(\mathrm{Tr}\,A_a\), parametrized by a symmetric matrix \(\Lambda_{ab}\):
\begin{equation}
S_{\rm tr}=\frac{1}{4\pi}\sum_{a,b}\Lambda_{ab}
\int\Tr A_a\wedge d\Tr A_b,\qquad \Lambda^T=\Lambda.
\label{eq:em-cs-action}
\end{equation}
In the symmetric determinant regularization, parity-anomaly cancellation requires $K_{AB}+\tfrac12\sum_\Phi\sum_{\rho\in R_\Phi}\rho_A\rho_B\in\mathbb Z$, where $A=(a,i)$ labels a Cartan generator. The matter contribution is $3N$ on the diagonal and $-3/2$ between distinct nodes. Thus $\Lambda_{aa}\in\mathbb Z$ and $\Lambda_{a\ne b}\in\mathbb Z+\tfrac12$ \cite{Benini:2011cma}.

We further impose \(\Lambda e=0\), with \(e=(1,1,1)^T\), so that the mixed CS terms induce no additional gauge charge on monopoles carrying identical fluxes at all three nodes. This is independent of anomaly cancellation. The allowed matrices are \cite{Gang:2011jj}
\begin{equation}
\Lambda=\frac12\begin{pmatrix}
2l_1&-l_1-l_2+l_3&-l_1+l_2-l_3\\
-l_1-l_2+l_3&2l_2&l_1-l_2-l_3\\
-l_1+l_2-l_3&l_1-l_2-l_3&2l_3
\end{pmatrix},
\label{eq:em-lambda}
\end{equation}
with integer $l_a$ and odd $\sum_a l_a$. Define $p=(2,-1,-1)^T$, $v=(0,-1,1)^T$ and
\begin{equation}
D \equiv v^T\Lambda v =-l_1+2l_2+2l_3,\quad d=l_2-l_3,\quad l_0=l_1-D/3.
\label{eq:em-parameters}
\end{equation}
Writing $\Lambda_*=\Lambda(1,1,1)$ gives
\begin{equation}
\begin{aligned}
\Lambda&= \Lambda_{\rm ref} +\frac{l_0}{4}pp^T
                    +\frac d4(pv^T+vp^T) .
\end{aligned}
\label{eq:em-decomposition}
\end{equation}
Thus $D$ measures the coupling in the relative direction $v$, while $d$ mixes the two relative directions. The main text uses the physical representative $3\Lambda_*$ at $D=9$. The reference matrix \( \Lambda_{\rm ref} \equiv (D/3)\Lambda_*\) is anomaly-admissible only for odd integer $D/3$; at $D=0$ we instead use $\Lambda_0=\Lambda(2,1,0)$.

\heading{Why only $D$ matters?}
Denote $S_a=\sum_{i=1}^{N}u_i^{(a)}$. At fixed sources, the product of the Bethe equations \eqref{eq:bae} gives $kp^TS=2\pi\widehat m$, with $\widehat m\in\mathbb Z$. We consider regular Bethe roots for which the phases \(u_i^{(a)}\) can be chosen so that \(p^TS=0\). For every physical root this is true at $k=1$: shifting one node-2 holonomy by $2\pi\widehat m$ changes no fugacity. At integer $k>1$, it is available only in the class $\widehat m=0\pmod k$.

For a reference root $u$ at $\Lambda_{\rm ref}$, the corresponding root at $\Lambda$ is
\begin{equation}
u_i^{\prime(a)}=u_i^{(a)}+z_0,\qquad
z_0=-\frac{d}{4k}(S_3-S_2).
\label{eq:em-root-map}
\end{equation}
This is because, on $p^TS=0$, Eq.~\eqref{eq:em-decomposition} changes the trace force by $(\Lambda - \Lambda_{\rm ref})S = d(v^TS)p/4$. The ordinary CS force changes by $kz_0p$, and the cancellation fixes $z_0$ above. All matter arguments remain unchanged. Consequently $l_0$ leaves the roots unchanged and $d$ translates their common center.

The shift in Eq.~\eqref{eq:em-root-map} also preserves the Jacobian determinant. To see why, use $h=2S_1-S_2-S_3$ as one coordinate and replace one logarithmic Bethe equation by the sum of all of them. In this sum the matter terms cancel pairwise, and the mixed CS terms cancel because $\Lambda e=0$. So fixed sources, the sum of the logarithmic Bethe equations obeys \(\delta[-i\sum_{a,i}\log\mathcal B_i^{(a)}]=k\,\delta h\). Expanding the determinant along the row of $h$ leaves the same factor $k$ in both theories and a determinant computed at fixed $h$.

We can fix $h=0$, Eq.~\eqref{eq:em-root-map} identifies the remaining equations. In coordinates given by the mean holonomy and independent holonomy differences, the map shifts the mean by $z_0$ and leaves all differences fixed. Since $z_0$ depends only on the difference $S_3-S_2$, this change of variables has unit determinant. Thus at corresponding roots,
\begin{equation}
\det J_\Lambda(u')=\det J_{\rm ref}(u).
\label{eq:em-shear}
\end{equation}
The vector and matter factors depend only on holonomy differences. Each $P_a=e^{\ii S_a}$ acquires a factor $e^{\ii Nz_0}$, so the topological flux insertion gives
\begin{equation}
\frac{Z^\Lambda_{u'}}{Z^{\rm ref}_u}
=\exp\!\left(\ii Nz_0\sum_a\ttop_a\right).
\label{eq:em-weight}
\end{equation}
Corresponding single-vacuum weights therefore agree when $\sum_a\ttop_a=0$, at any genus with matching chiral fluxes. 

\heading{Logarithmic corrections at $D=9$}
At $k=1$, consider the auxiliary path $\Lambda(D)=\Lambda_0+(D/4)vv^T$. Set $w=\ii u$ and $R=\sum_i(w_i^{(2)}-w_i^{(3)})$. The added CS term shifts the logarithmic Bethe equations at nodes 2 and 3 by $+DR/4$ and $-DR/4$, respectively. Adding opposite sources $+\zeta$ and $-\zeta$ to the $\Lambda_0$ equations at these nodes gives the same root when $\zeta=DR/4$. Varying $\zeta$ along a branch of the source equations defines $\chi=-dR/d\zeta$. If their Jacobian $J_{\rm src}$ at fixed $\zeta$ is invertible, the rank-one determinant identity gives
\begin{equation}
\det J(D)=\det J_{\rm src}\left(1+\frac{D\chi}{4}\right).
\label{eq:em-collective}
\end{equation}
This identity is exact. For a genus-zero residue it contributes $\log|1+D\chi/4|$ to $F$. We numerically find that
\begin{equation}
\chi=-\frac49+\chi_1N^{-1/2}+o(N^{-1/2}),\qquad \chi_1\ne0,
\label{eq:em-response}
\end{equation}
then the constant part $1-D/9$ cancels at $D=9$ and this contribution becomes $-\tfrac12\log N+O(1)$. This explains the correct logarithmic coefficient at $D = 9$. 

\end{document}